\documentclass[prl,reprint,showpacs] {revtex4-1}
\usepackage{amsmath,graphicx}

\usepackage{times}
\usepackage[colorlinks=true,linkcolor=blue,urlcolor=blue,citecolor=blue]{hyperref}

\def \dif {{d}}

\date\today

\begin{document}

\title{Dynamics of a Mobile Impurity in a One-Dimensional Bose Liquid}

\author{Aleksandra Petkovi\' c}
\author{Zoran Ristivojevic}
\affiliation{Laboratoire de Physique Th\'{e}orique, Universit\'{e} de Toulouse, CNRS, UPS, 31062 Toulouse, France}

\begin{abstract}

We develop a microscopic theory of a quantum impurity propagating in a one-dimensional Bose liquid. As a result of scattering off thermally excited quasiparticles, the impurity experiences the friction. We find that, at low temperatures, the resulting force scales either as the fourth or the eighth power of temperature, depending on the system parameters. For temperatures higher than the chemical potential of the Bose liquid, the friction force is a linear function of temperature. Our approach enables us to find the friction force in the crossover region between the two limiting cases. In the integrable case, corresponding to the Yang-Gaudin model, the impurity becomes transparent for quasiparticles and thus the friction force is absent. Our results could be further generalized to study other kinetic phenomena.
\end{abstract}

\maketitle

The motion of distinguishable particles (impurities) through the Bose liquid is an old subject relevant for a large class of systems. One of the first examples is the dynamics of $^3${He} atoms in a $^4${He} superfluid, studied by Landau and co-workers   \cite{landau+49,khalatnikov2000introduction}. The resurgent interest in this topic has arisen due to various realizations with cold atoms \cite{chikkatur2000suppression, palzer2009quantum,zipkes2010trapped,schmid2010dynamics,weitenberg2011single-spin,catani2012quantum,fukuhara2013quantum} that, e.g., could simulate the motion of a spin excitation in a magnetic system. The achieved unprecedented experimental control of the system urges us to develop refined methods that accurately describe and predict fundamental physical phenomena. The studies of mobile impurities in one-dimensional liquid environments have contributed to our understanding of the excitation spectrum and the effective mass \cite{mcguire1965interacting,yang1967exact,gaudin1967systeme,fuchs2005spin,girardeau2009motion,lamacraft2009dispersion}, various response functions \cite{castella1993exact,zvonarev2007spin,matveev2008spectral,kamenev2009dynamics,imambekov2012one-dimensional}, the impurity dynamics \cite{castro_neto1996dynamics,astrakharchik2004motion,gangardt2009bloch,matveev2012scattering,johnson2012breathing}, and nonequilibrium phenomena \cite{bonart2012nonequilibrium,peotta2013quantum,burovski2014momentum,knap2014quantum,robinson2016motion}.

A superfluid is the zero-temperature ground state of a quantum Bose liquid  \cite{Landau9}. A slow impurity flows through this system without dissipation since it cannot emit quasiparticles. The dissipation occurs at finite temperature due to scattering with thermally excited quasiparticles. The resulting friction force exerted on the impurity by a one-dimensional liquid scales as the fourth power of temperature \cite{castro_neto1996dynamics,Note1}. 
However, the detailed analytical study of the impurity dynamics is challenging as it requires a careful treatment of the correlation effects in the liquid in the presence of a mobile particle. 

Recently, considerable progress in this direction has been made using the so-called mobile impurity formalism \cite{pustilnik2006dynamic,pereira2008exact,imambekov2009phenomenology,gangardt2009bloch,imambekov2012one-dimensional,matveev2012scattering,schecter2012dynamics}. In this phenomenological approach, one considers the quasiparticle excitations of the longest wavelengths, larger than the de Broglie wavelength of the impurity. One thus accounts for scattering processes where the impurity experiences a small relative change of the momentum. The approximation of the linear quasiparticle spectrum does not appear to be a limitation of this approach in describing the scattering at low temperatures, which only excite  the quasiparticles at small momenta. For the friction force exerted on the impurity, one confirms the above-mentioned scaling obtained in Ref.~\cite{castro_neto1996dynamics}, finding the full dependence on the system parameters \cite{gangardt2009bloch,schecter2012dynamics,matveev2012scattering}.

In this Letter we study the dynamics of an impurity in a one-dimensional Bose liquid. We find that by controlling the system parameters, the friction force can dramatically change its temperature dependence from $T^4$ \cite{castro_neto1996dynamics,gangardt2009bloch,matveev2012scattering} to $T^8$. Contrary to the naive expectation, this result, albeit valid at low temperatures, requires us to account for the nonlinearity of the quasiparticle spectrum at low momenta. We develop a microscopic hydrodynamic approach that considers the nonlinear quasiparticle spectrum and thus enables us to analytically calculate the full dependence of the friction force in a wide range of temperatures. For temperatures above the chemical potential of the Bose liquid, we find linear $T$ dependence. 

Apart from being relevant to experiments, our results bring new insights into various phenomena. The detailed knowledge of the friction force on the impurity is necessary for studies of the driven impurity dynamics as well as for the understanding and detection of Bloch oscillations \cite{gangardt2009bloch,schecter2012dynamics}. The Casimir interaction between impurities immersed in the quantum liquid \cite{casimir1948,kardar1999``friction,schecter2014phonon-mediated} will show new regimes \cite{unpub} since it is governed by the same scattering mechanism as the friction force. Finally, the developed microscopic theory allows us to fully characterize the stochastic motion of the impurity caused by collisions with the host liquid by studying the corresponding kinetic equation \cite{unpub}.

The system of one-dimensional interacting bosons is described by the Hamiltonian \cite{lieb1963exact}
\begin{align}\label{HL}
H_L=\int\dif x\left[\frac{\hbar^2}{2m}(\nabla\psi^\dagger)(\nabla\psi)+ \frac{g}{2}n^2\right].
\end{align}
Here, $\psi(x)$ and $\psi^\dagger(x)$ are the bosonic single particle operators that satisfy the standard commutation relation $[\psi(x),\psi^\dagger(y)]=\delta(x-y)$, while $n(x)=\psi^\dagger(x)\psi(x)$ denotes the density of the particles. By $m$ we denote the mass of the particles, while $g$ is the strength of the short-range repulsion. The minimal model that captures the impurity dynamics  is  
\begin{gather}
\label{H}
H=H_L+\frac{\hbar^2}{2M}\int\dif x (\nabla\Psi^\dagger)(\nabla\Psi)
+G\int\dif x\, \Psi^\dagger\Psi n.
\end{gather}
The second term of Eq.~(\ref{H}) is the kinetic energy of the impurity of the mass $M$, described in terms of the field operator $\Psi(x)$. The impurity interacts with the liquid via a short-range density-density interaction of the strength $G$, as given by the last term in Eq.~(\ref{H}). We consider the case of weak interaction, when the Luttinger liquid parameter $K=\pi \hbar \sqrt{n_0/mg}\gg 1$. Here, $n_0$ is the mean density of the bosons. We study the impurity weakly coupled to the liquid, $G\ll g\sqrt{K}$.

We transform the Hamiltonian of the liquid by representing the bosonic field operator as \cite{popov1972theory,haldane1981effective}
$\psi^\dagger=\sqrt n\,e^{i\theta}$, where the density $n$ and the phase $\theta$ satisfy the standard bosonic commutation relation $[n(x),\theta(y)]=-i\delta(x-y)$. Equation (\ref{HL}) then takes the form \cite{popov1972theory}
\begin{align}\label{HL1}
H_{L}=\int\dif x\left[\frac{\hbar^2 n}{2m}(\nabla\theta)^2+\frac{\hbar^2(\nabla n)^2}{8mn}+\frac{g}{2}n^2\right].
\end{align}
To describe the low energy excitations of the liquid, we reexpress the bosonic density operator as  \cite{haldane1981effective,cazalilla2011one}
\begin{align}\label{density}
n=n_0+\nabla\varphi/\pi.
\end{align}
Here, $\varphi$ satisfies the commutation relation $[\nabla\varphi(x),\theta(y)]=-i\pi\delta(x-y)$. The field $\nabla\varphi$ controls the fluctuations of the density, which are small for the excitations of wave vectors smaller than $n_0$ \cite{Note2}. 
Upon substituting Eq.~(\ref{density}) into Eq.~(\ref{HL1}) we can expand the Hamiltonian $H_L$ in powers of $\nabla\varphi$. We then diagonalize the quadratic part of $H_L$  by using the normal mode expansion, which connects the bosonic fields $\varphi$ and $\theta$ with the bosonic quasiparticle operators $b_p$ and $b_p^\dagger$:
\begin{gather}\label{nablaphi}
\nabla\varphi(x)=\sum_p \sqrt{\frac{\pi^2 n_0}{2Lm\varepsilon_p}}|p| e^{i p x/\hbar} (b_{-p}^\dagger+b_p),\\
\label{nablatheta}
\nabla\theta(x)=\sum_p \sqrt{\frac{m \varepsilon_p}{2 L\hbar^2 n_0}}\,\text{sgn}(p)e^{i p x/\hbar} (b^\dagger_{-p}-b_{p}).
\end{gather}
Here, $L$ denotes the system size. Taking into account the second term in Eq.~(\ref{H}), which we transform by introducing $\Psi(x)=\frac{1}{\sqrt{L}}\sum_P e^{i P x/\hbar}B_P$, we eventually obtain the diagonal form of the quadratic part of the total Hamiltonian (\ref{H}):
\begin{align}\label{H0diag}
H_0=\sum_p\varepsilon_p b_p^\dagger b_p+\sum_P E_P B^\dagger_P B_P.
\end{align}
As expected, it is a sum of decoupled excitations of the liquid and the impurity. In Eq.~(\ref{H0diag}), the excitation spectrum of the liquid is given by the Bogoliubov dispersion relation \cite{kulish1976comparison} $\varepsilon_p=\sqrt{v^2 p^2+{p^4}/{4 m^2}}$, where $v=\sqrt{gn_0/m}$ denotes the sound velocity. We emphasize that our approach accounts for the excitations of the liquid at wave vectors below $n_0$. It includes the full crossover between the linear and quadratic regimes, separated by the characteristic momentum $mv$. The dispersion of the impurity in Eq.~(\ref{H0diag}) is $E_P=P^2/2M$.

The impurity is coupled to the liquid as it can emit and absorb its quasiparticles. Such processes are described by the last term of Eq.~(\ref{H}). In normal modes, it takes the form
\begin{align}\label{V1nm}
V_1=\sum_{P_1,P_2,p} \Gamma(p) B^\dagger_{P_1}B_{P_2}(b^\dagger_{-p}+b_p)
 \delta_{P_1,P_2+p}.
\end{align}
Here, $\Gamma(p)=G\sqrt{{n_0 p^2}/{2Lm\varepsilon_p}}$ describes the coupling between the two subsystems. In Eq.~(\ref{V1nm}) we omitted an additive constant, $Gn_0$. 

To develop a consistent theory of the dynamics of a mobile impurity in one dimension, one should be careful and, in addition to $V_1$, must consider the leading correction to the quadratic Hamiltonian of the liquid. Such a term describes the residual interaction of Bogoliubov excitations. It arises from the expansion of Eq.~(\ref{HL}) in the cubic terms in $\nabla\varphi$ and $\nabla\theta$. In the normal mode representation, the cubic perturbation consists of two types of terms. The term that describes the processes where three Bogoliubov quasiparticles are either created or annihilated turns out not to be important for our purposes. Here, we need the remaining part of the cubic perturbation that takes the form
\begin{align}\label{V3}
V_3={}&\frac{v^2}{\sqrt{32Ln_0 m}} \sum_{p_1,p_2,p_3}f\left(p_1,p_2,p_3\right)\frac{|p_1p_2p_3|} {\sqrt{\varepsilon_{p_1}\varepsilon_{p_2}\varepsilon_{p_3}}} \notag\\
&\times  (b^\dagger_{p_1} b^\dagger_{p_2}b_{-p_3} + \text{H.c.}) \delta_{p_1+p_2+p_3,0}.
\end{align}
In Eq.~(\ref{V3}) we introduced the dimensionless function
\begin{align}\label{fpm}
f(p_1,p_2,p_3)={}&\frac{1}{v^2}\left( \frac{\varepsilon_{p_1}\varepsilon_{p_2}} {p_1p_2} - \frac{\varepsilon_{p_1}\varepsilon_{p_3}} {p_1p_3} -\frac{\varepsilon_{p_2}\varepsilon_{p_3}} {p_2p_3}\right)\notag\\
&+\frac{1}{4m^2 v^2}(p_1p_2+p_1p_3+p_2p_3).
\end{align}

A simple kinematic argument shows that an impurity slower than the sound velocity $v$ cannot relax by emitting Bogoliubov quasiparticles \cite{Landau9}. However, it relaxes due to collisions with thermally excited quasiparticles. The main process involves impurity scattering off a quasiparticle that is absorbed while another one is emitted. The corresponding matrix element has two contributions in second order perturbation theory: (i) one arises from $V_1$ [Eq.~(\ref{V1nm})],  while (ii) the other involves both perturbations, $V_1$  and $V_3$ [Eq.~(\ref{V3})]. Let us estimate the two terms (i) and (ii). Perturbations (\ref{V1nm}) and (\ref{V3}) both involve the momentum of a Bogoliubov quasiparticle, which has the characteristic value $mv$. For $V_1$, we then find the typical energy scale that is of the order of $Gn_0/\sqrt{K}$. The corresponding energy scale for $V_3$ is $gn_0/\sqrt{K}$. Through the denominators appearing in  second order perturbation theory, the masses get involved. For slow impurity one finds that contributions (i) and (ii) are both of the same order provided that $G/M$ is of the order of $g/m$. Thus, although the perturbation (\ref{V3}) does not involve the impurity operator, for the problem of impurity dynamics it must be treated on equal footing as the perturbation (\ref{V1nm}).

Denoting by $P$ ($p$) the momentum of the impurity (quasiparticle) in the initial state and by the primed symbols the corresponding momenta in the final state, the scattering matrix element in terms of the $T$ matrix is given by $t_{P,p}^{P',p'}=\langle B_{P'}b_{p'}|T|B_P^\dagger b_p^\dagger\rangle$. Its leading contribution on the mass shell is 
\begin{align}\label{t}
t={}&\frac{\Gamma(p)\Gamma(p')}{E_{P}-E_{P+p}+\varepsilon_{p}}+ \frac{\Gamma(p)\Gamma(p')}{E_{P}-E_{P-p'}-\varepsilon_{p'}}  +\frac{v^2 \Gamma(p-p')}{\sqrt{8Ln_0 m}}\notag\\
&\times\frac{|p p' (p-p')|}{ \sqrt{\varepsilon_{p} \varepsilon_{p'} \varepsilon_{p-p'}}}\biggl[\frac{f(p-p',p',-p)}{\varepsilon_{p}-\varepsilon_{p'}-\varepsilon_{p-p'}}+ \frac{f(p'-p,p,-p')}{\varepsilon_{p'}-\varepsilon_{p}-\varepsilon_{p-p'}}\biggr]
\end{align}
where, for easy notation, we introduced $t_{P,p}^{P',p'}=t\, \delta_{P+p,P'+p'}$. Here, the Kronecker delta accounts for the momentum conservation.

The matrix element for the impurity scattering off quasiparticles, Eq.~(\ref{t}), is the central object that determines the dynamic characteristics of the impurity. In the following, we focus on the friction force exerted on the impurity by the liquid. Using Fermi's golden rule it can be expressed as
\begin{align}\label{force}
F={}&\frac{2\pi}{\hbar}\sum_{P',p,p'} |t_{P,p}^{P',p'}|^2 (P'-P) n_p(1+n_{p'})\notag\\&\times  \delta(E_{P}+\varepsilon_{p}-E_{P'}-\varepsilon_{p'}).
\end{align}
Here, $n_p=(e^{\varepsilon_p/T}-1)^{-1}$ is the Bose occupation factor.

We begin our analysis of Eq.~(\ref{force}) by considering the regime of temperatures below the chemical potential of the Bose gas, $T\ll \text{min}(m,M)v^2$. In this case, the occupation factor $n_p$ is appreciable only at momenta $p\sim T/v\ll mv$. The conservation laws of the momentum and energy then impose a small change of the impurity momentum, enabling us to expand the matrix element (\ref{t}). At the lowest order in $\delta P=P'-P$, we find
\begin{align}\label{eq:t0}
t=\frac{G}{4L} \left(1-\frac{Gm}{gM}\right) \frac{|\delta P|}{m\sqrt{v^2-V^2}}.
\end{align} 
Here, we have introduced the impurity velocity $V=P/M$.
In Eq.~(\ref{eq:t0}) (and the forthcoming ones for $t$ in other regimes) the momenta of the quasiparticles are expressed as functions of the impurity initial and final momentum using the conservation laws. Substituting the matrix element (\ref{eq:t0}) into the expression (\ref{force}), we obtain the friction force at low temperatures \cite{matveev2012scattering},
\begin{align}\label{eq:FT0}
F=-\frac{2\pi^3}{15}\left(1-\frac{Gm}{gM}\right)^2\frac{G^2 T^4}{\hbar^3 m^2}\frac{V(v^2+V^2)}{(v^2-V^2)^5}.
\end{align}
When the impurity momentum $P$ is in the very near vicinity of $M v$, perturbation theory breaks down. This occurs because there the renormalized impurity dispersion  begins to significantly deviate from $E_P$, as can be shown by considering the correction to the impurity dispersion due to its coupling to the liquid.

In the special case with $m=M$ and $g=G$, the Hamiltonian (\ref{H}) corresponds to the Yang-Gaudin model \cite{yang1967exact,gaudin1967systeme}, which is integrable. Thus, one should expect the absence of scattering of the impurity off quasiparticles. Indeed, we find that the matrix element (\ref{t}), and thus the force (\ref{force}), nullifies in this case when all of the contributions in Eq.~(\ref{t}) are taken into account. However, we notice that the expanded matrix element (\ref{eq:t0}) nullifies when a less restrictive condition $G/g=M/m\neq1$ is satisfied. In this case, we must expand Eq.~(\ref{t}) to account for higher order contributions in $\delta P$. It yields 
\begin{align}
\label{eq:t0spec}
t=\frac{G}{64L}\left(1-\frac{m^2}{M^2}\right)\frac{\left|\delta P\right|^3}{m^3 v^2 \sqrt{v^2-V^2}},
\end{align}
which then leads to 
\begin{align}\label{lowTspec}
F={}&-\frac{2\pi^7}{15}\left(1-\frac{m^2}{M^2}\right)^2\frac{G^2 T^8}{ \hbar^3 m^6 v^4} \frac{V(v^2+V^2)}{(v^2-V^2)^9}\notag\\
&\times (v^4+6V^2v^2+V^4).
\end{align}
This result behaves as the eighth power of temperature, as opposed to the fourth power that appears in Eq.~(\ref{eq:FT0}). We emphasize that Eq.~(\ref{lowTspec}) is the dominant contribution in the force for a range of parameters $|1-Gm/gM|<|1-m^2/M^2|(\pi T/mv^2)^2$ for slow impurity. Interestingly, although the force (\ref{lowTspec}) is valid at low temperatures, its derivation requires the nonlinearity of the quasiparticle spectrum, contained in the Bogoliubov dispersion. Hence, it is important to account for the second term in Eq.~(\ref{HL1}), which is known as the quantum pressure. This should be contrasted with the result (\ref{eq:FT0}), for which is  sufficient only to know  the linear part of the spectrum.

\begin{figure}
\includegraphics[width=\columnwidth]{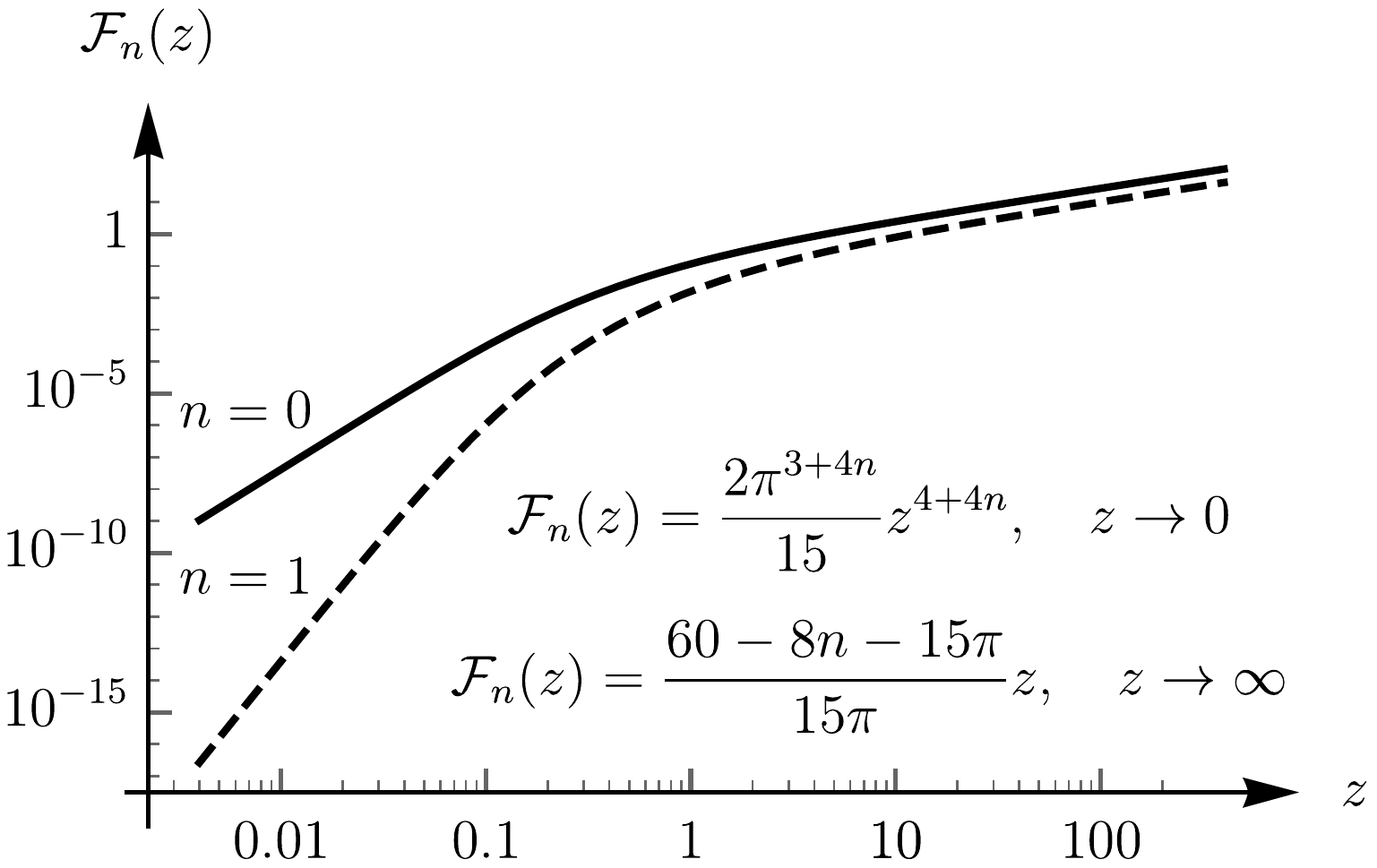}
\caption{Plot of the crossover function $\mathcal{F}_n(z)$ given by Eq.~(\ref{crossover}). The solid (dashed) line represents the case $n=0$ ($n=1$).}\label{fig1}
\end{figure}

Next we consider the regime of heavy impurity $M\gg m$, which enables us to analytically explore the friction force in a wide range of temperatures. We evaluate the matrix element (\ref{t}) and find the leading two contributions (in a small $m/M$): 
\begin{align}\label{eq:tlargeM}
t=\frac{G}{L}\frac{|\delta P|\left[16 m^2 v^2 \left(1-{Gm}/{gM}\right)+\delta P^2\right] }{(16 m^2 v^2+ \delta P^2)^{3/2}}.
\end{align}
In a case with comparable coupling constants, we can neglect the subleading term in the previous equation. Introducing the dimensionless crossover function
\begin{align}
\label{crossover}
\mathcal{F}_n(z)=\frac{1}{2\pi z} \int_0^\infty{}{}  \frac{d\epsilon\left(\sqrt{1+\epsilon^2}-1\right)^{4n+2}} {\sinh^2\left(\frac{\epsilon}{2z}\right) \epsilon^{4n}(1+\epsilon^2)},
\end{align}
(see Fig.~1), the friction force can be expressed as 
\begin{align}\label{eq:largeM}
F&=-\frac{ G^2 m^2 V}{\hbar^3}\mathcal{F}_0\left(\frac{T}{m v^2}\right).
\end{align}
In the regime of low temperatures, $T\ll mv^2$, the latter result can be further simplified to $F=-{2\pi^3}{G^2 T^4 V}/{(15 \hbar^3 m^2 v^8)}$, which is
in agreement with Eq.~(\ref{eq:FT0}) at low $V$. In the regime of high temperatures, $T\gg mv^2$, we find the linear $T$ dependence:
\begin{align}\label{FhighT}
F=-\frac{4-\pi}{\pi}\frac{G^2 m T V}{\hbar^3v^2}.
\end{align}
The friction force in the crossover regime $T\sim mv^2$ is given by Eq.~(\ref{eq:largeM}) and is expressed in terms of the universal function of temperature $\mathcal{F}_0(T/mv^2)$ given by Eq.~(\ref{crossover}). We note that Eq.~(\ref{eq:largeM}) applies to both cases of repulsive and attractive interaction $G$.

The case $G/g=M/m (\gg 1)$ requires special care. As shown in Eq.~(\ref{lowTspec}), under this condition the friction force at low temperatures behaves as $T^8$, contrary to the usual $T^4$ dependence  \cite{castro_neto1996dynamics,schecter2012dynamics,matveev2012scattering} in Eq.~(\ref{eq:FT0}). At high temperatures it turns out that only the numerical prefactor differs from the one in Eq.~(\ref{FhighT}). The reason is that unlike the low temperature regime, where the change of the impurity momentum $\delta P$ is controlled by the temperature, at high temperatures it is not and becomes of the order of $mv$. Therefore, even though the dependence on $\delta P$ of the matrix element (\ref{eq:tlargeM}) changes  when setting $M/m=G/g$, the resulting friction force (\ref{force}) is not considerably influenced. Its full temperature dependence is given by
\begin{align}\label{eq:crossoverspec}
F=-\frac{G^2 m^2 V}{\hbar^3}\mathcal{F}_1\left(\frac{T}{m v^2}\right),
\end{align}
where the function $\mathcal{F}_1$ takes the form (\ref{crossover}). We evaluate this expression at low temperatures and obtain
$
F=-{2\pi^7}{G^2T^8V}/{15\hbar^3 m^6 v^{16}}, 
$
which is in agreement with Eq.~(\ref{lowTspec}). At high temperatures, the force (\ref{eq:crossoverspec}) becomes
$
F=-{(52-15 \pi)}{G^2 m T V}/{{15 \pi}\hbar^3 v^2}.
$

The results (\ref{lowTspec}) and (\ref{eq:crossoverspec}) cease to be valid at extremely low temperatures $T\ll mv^2/\sqrt{K}$, with $K\gg 1$. In this regime the quasiparticles of very low momenta are predominantly excited. They cannot be described by the Bogoliubov dispersion \cite{imambekov2012one-dimensional}, and the scattering problem requires a separate study. On the other hand, this limitation does not apply to the results (\ref{eq:FT0}) and (\ref{eq:largeM}), and they are valid even at $T\to 0$. We note that Eqs.~(\ref{eq:largeM}) and (\ref{eq:crossoverspec}) assume not too high temperatures, $T\ll M v^2$. The latter condition for a very heavy impurity is replaced by $T\ll\hbar^2 n_0^2/M$, while the impurity momentum should be less than $\pi\hbar n_0$.

To conclude, we have studied the scattering problem of a mobile impurity in a one-dimensional weakly repulsive Bose gas.  We have developed the microscopic description that enabled us to analytically calculate the friction force exerted on the impurity in a wide range of temperatures. The main results of the Letter are given by Eqs.~(\ref{t}), (\ref{lowTspec}), (\ref{eq:largeM}), and (\ref{eq:crossoverspec}). Finally, the experimental realizations with ultracold gases provide an ideal playground where our results could be tested, since impurities can be realized, e.g., by mixing different atoms, while the interactions can be tuned using the Feshbach resonance \cite{chin2010feshbach}. We emphasize that our approach, particularly the result for the scattering matrix element (\ref{t}), can be used to study other kinetic phenomena.


%

\end{document}